\documentclass[aps,reprint,prx,superscriptaddress,floats,floatfix,nobibnotes]{revtex4-2}

\usepackage{physics}
\usepackage{xcolor}
\usepackage{latexsym}
\usepackage{amsmath}
\usepackage{graphicx}

\usepackage[colorlinks=true,citecolor=blue,linkcolor=blue,urlcolor=blue]{hyperref}
\usepackage{times}
\usepackage{cleveref}
\usepackage{ragged2e}
\usepackage{orcidlink}

\newcommand{\UFSCar}{Departamento de Física, Universidade Federal de São Carlos, Rodovia Washington Luís, km 235 - SP-310, 13565-905 São Carlos, SP, Brazil}
\newcommand{\Unesp}{Universidade Estadual Paulista (UNESP), Instituto de Ciências e Engenharia, 18409-010 Itapeva, São Paulo, Brazil}

\begin{document}

\title{Quantum phase estimation for nondestructive monitoring and Wigner tomography of bosonic fields}

\author{Lucas R. S. Santos~\orcidlink{0000-0001-9700-884X}}
\email[]{lucasrss@estudante.ufscar.br}
\affiliation{\UFSCar}

\author{Ciro M.~Diniz~\orcidlink{0000-0002-7602-0468}}
\address{\UFSCar}

\author{Daniel Z. Rossatto~\orcidlink{0000-0001-9432-1603}}
\address{\Unesp}

\author{Celso J. Villas-Boas~\orcidlink{0000-0001-5622-786X}}
\affiliation{\UFSCar}

\date{\today}

\begin{abstract}
    Quantum phase estimation is usually introduced as an algorithmic primitive for extracting eigenphases of unitary operators. Here we show that, when implemented through a dispersive light-matter interaction, it can also be used as a nondestructive measurement tool for bosonic fields. We consider a bosonic mode coupled to a multi-qubit register and calibrate the photon-number dependent phase shifts so that the register performs a number-resolved quantum phase estimation readout. Repeating this readout during dissipative evolution enables nondestructive monitoring of photon-number dynamics. We then show that the same readout can be converted into a Wigner-tomography reconstruction by applying phase-space displacements before the quantum phase estimation block. Numerical reconstructions for Fock, coherent, and even/odd Schrödinger-cat states show the expected nonclassical phase-space structures and near-unity Wigner-overlap fidelities. The protocol provides a unified route to nondestructive monitoring and state tomography of bosonic fields, with direct relevance for bosonic-state characterization, calibration, and control in superconducting quantum architectures.
\end{abstract}

\maketitle

\section{Introduction}

Quantum non-demolition (QND) measurements play a central role in quantum optics and quantum information because they provide a route to extract information about an observable while, ideally, preserving the physical degree of freedom being monitored~\cite{grangier1998qnd, haroche2006exploring, Guerlin2007}. In the context of bosonic systems, this idea is especially appealing: rather than destroying the field through an absorptive measurement, one aims to infer quantities such as the photon number by coupling the field dispersively to an ancillary system and transferring the information to a readable probe~\cite{grangier1998qnd,Guerlin2007,deleglise2008reconstruction}. This capability is not only of foundational interest. It is also a practical requirement whenever one wishes to monitor stochastic trajectories of open quantum systems, observe quantum jumps, perform feedback and stabilization, or repeatedly interrogate a field without re-preparing it after every measurement round~\cite{Guerlin2007, deleglise2008reconstruction, haroche2006exploring, sun2014tracking}.

These ideas found one of their clearest experimental realizations in cavity quantum electrodynamics (cavity-QED), particularly in the work of Ref.~\cite{Guerlin2007}. There, the photon number stored in a high-$Q$ microwave cavity is inferred from the dispersive phase shifts accumulated by Rydberg atoms crossing the cavity~\cite{lutterbach1997direct, Guerlin2007}. By measuring long sequences of such atoms, it becomes possible to observe the progressive collapse of an initially broad photon-number distribution toward a Fock state, and then to follow the subsequent staircase-like decay of the field energy through quantum jumps. These experiments established cavity-QED as a consolidated platform for QND monitoring of bosonic fields, showing in practice that repeated, nondestructive interrogation can reveal the real-time dynamics of relaxation and measurement-induced collapse~\cite{Guerlin2007, haroche2006exploring}.

Beyond cavity-QED, related nondestructive monitoring ideas have also been explored in trapped-ion and superconducting platforms. In trapped ions, QND schemes for the vibrational energy showed that the motional excitation of the ion can be accessed without demolishing the vibrational state itself~\cite{harrison1997qndion}. In a complementary direction, Ref.~\cite{Travaglione2001} shows that the quantum phase estimation (QPE) algorithm can project a system onto eigenstates of an operator of interest, suggesting that phase-estimation routines may also be viewed as physically meaningful tools for selective measurements of quantized systems. In superconducting circuits, repeated QND parity measurements of microwave cavity fields have enabled real-time tracking of photon jumps, demonstrating that fast and repeated bosonic-state monitoring is compatible with circuit-based quantum hardware~\cite{sun2014tracking}.

However, photon-number monitoring alone provides only partial information about the quantum state of the field. Even when repeated QND measurements reveal the dynamics of the field energy, they do not characterize phase-space coherences, interference fringes, or other signatures of nonclassicality. This makes it difficult to identify and distinguish general nonclassical states. By contrast, the Wigner function provides a more complete phase-space portrait, revealing both the quasiprobability structure of the field and genuinely quantum features such as negativities and interference patterns~\cite{lutterbach1997direct, nogues2000measurement, bertet2002direct, deleglise2008reconstruction}. In this context, a nondestructive, number-resolved measurement scheme capable of accessing both photon-number dynamics and, after suitable phase-space displacements, the Wigner function would be particularly valuable.

This observation leads to the central question addressed in this work: can the same nondestructive, number-resolved measurement primitive be used both to monitor photon-number dynamics and, after suitable phase-space displacements, to reconstruct the Wigner function of the field? Addressing both tasks within a common readout architecture is particularly relevant for bosonic quantum information processing. In many descriptions of quantum algorithms, attention is often concentrated on the sequence of unitary gates that implements the desired evolution, while state preparation and readout are treated as separate implementation layers~\cite{nielsen2010quantum}. A more careful resource analysis, however, must also include the number of state preparations, measurement settings, and repetitions required to infer observables or reconstruct the output state~\cite{cardoso2021complexity}. Whenever the output contains nontrivial coherences in the computational basis or in a bosonic mode, extracting this information may require many identically prepared copies and repeated measurements. Consequently, the practical cost of a quantum protocol may be dominated by its readout and tomography stages even when the unitary circuit implementing the target evolution has favorable complexity.

Over the last few decades, remarkable tomographic achievements have been reported across distinct experimental platforms. In trapped ions, reconstructions of motional density matrices and Wigner functions were obtained by applying coherent displacements with different amplitudes and phases, repeatedly probing the sideband dynamics, and using fluorescence detection to infer the displaced number-state populations~\cite{leibfried1996motional}. In optical continuous-variable systems, homodyne-based reconstructions of squeezed and nonclassical light required the collection of quadrature statistics over the optical phase, followed by a numerical reconstruction of the density matrix and Wigner function~\cite{breitenbach1997squeezed}. In cavity-QED, measurements of Wigner negativities and full reconstructions of coherent, Fock, and Schrödinger cat states required controlled field displacements and sequences of Ramsey probe atoms at many phase-space points and over repeated field realizations~\cite{nogues2000measurement, bertet2002direct, deleglise2008reconstruction}. In superconducting platforms, Wigner tomography of cavity cat states similarly combines calibrated cavity displacements, ancilla-assisted parity mapping, and repeated qubit readout over a phase-space grid~\cite{vlastakis2013cat}. Taken together, these results demonstrate the maturity and power of quantum-state tomography. At the same time, they highlight its resource demands: full reconstruction is inherently statistical and generally requires repeated state preparations, multiple measurement settings or phase-space points, and platform-specific control and readout sequences.

We address this question within a circuit-QED architecture in which a bosonic mode is dispersively coupled to a multi-qubit register. In this regime, the photon number of the resonator is encoded into phase shifts accumulated by the ancillary qubits, allowing the QPE algorithm~\cite{kitaev_1995} to be reinterpreted as a nondestructive, number-resolved readout tool for the field. The central point is that the QPE register provides access to the photon-number distribution rather than only to its parity. For the monitoring task, repeated QPE interrogations map photon-number information onto register outcomes while leaving the resonator energy unchanged, allowing photon-loss events to be tracked as discrete changes in the inferred photon number. For the tomography task, phase-space displacement operations are applied before the same QPE readout. Repeated measurements then provide the displaced photon-number distribution $P_n(\alpha)$ for each displacement amplitude $\alpha$, from which the displaced parity and therefore the Wigner function can be reconstructed. Thus, the same QPE-based measurement primitive provides dynamical photon-number information when applied sequentially and phase-space information when preceded by displacement operations. In a complementary application of the same dispersive QPE framework, Ref.~\cite{lucas2026largefockstates} combines photon-number encoding with quantum amplitude amplification to enable high-probability preparation of large Fock states. Together, these approaches illustrate how QPE-based dispersive interactions can be used not only for nondestructive readout and characterization, but also for bosonic-state preparation.

This article is organized as follows. Section~\ref{sec:model} introduces the physical model and derives the effective dispersive interaction underlying the protocol. Section~\ref{sec:protocol} presents the QPE readout and its implementation as a number-resolved, nondestructive measurement of the resonator field. Section~\ref{sec:qnd_monitoring_results} applies this readout to QND monitoring of dissipative photon-number dynamics and discusses the resulting quantum trajectories. Section~\ref{sec:tomography} extends the same QPE to Wigner tomography through controlled phase-space displacements and displaced-parity reconstruction. Finally, Sec.~\ref{sec:conclusion} summarizes the results and discusses perspectives for bosonic-state characterization and control in circuit-QED architectures.

\section{Physical model}\label{sec:model}

\begin{figure*}
    \centering
    \includegraphics[width = 1.0\textwidth, clip]{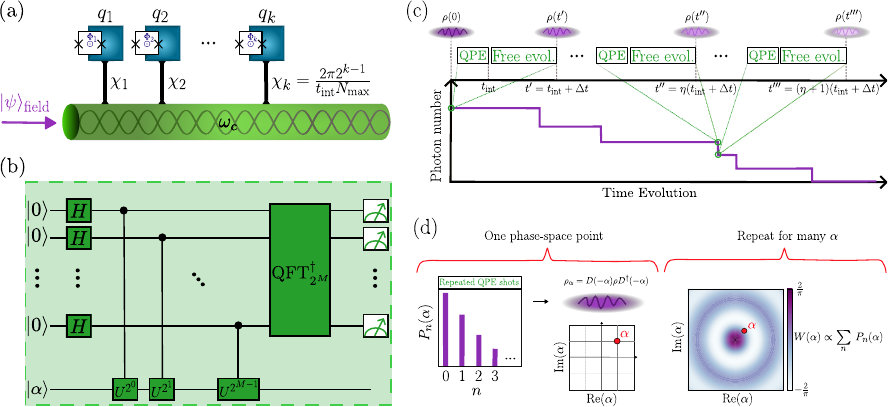}
    \caption{\justifying Circuit-QED protocol for QND monitoring and Wigner tomography of a bosonic field using quantum phase estimation (QPE).
    (a) Physical platform: a bosonic resonator mode at frequency $\omega_c$ is dispersively coupled to a register of $M$ frequency-tunable superconducting qubits. The flux-dependent detunings $\Delta_k(\Phi_k)$ control the dispersive shifts $\chi_k(\Phi_k)\simeq |g_k|^2/\Delta_k(\Phi_k)$, which are calibrated to implement the QPE controlled phases.
    (b) QPE readout: the register is prepared in a uniform superposition, acquires photon-number-dependent phases through the dispersive interaction, and is measured after the inverse quantum Fourier transform, yielding an outcome correlated with the resonator photon number.
    (c) QND monitoring: QPE interrogations are alternated with intervals of free cavity evolution, allowing photon-loss events to appear as discrete changes in the inferred photon number.
    (d) Wigner tomography: for each fixed phase-space displacement $\alpha$, the QPE readout is repeated to estimate the full displaced photon-number distribution $P_n(\alpha)$. Its parity sum gives one phase-space value, $W(\alpha)\propto\sum_n(-1)^nP_n(\alpha)$. Repeating the procedure over different $\alpha$ reconstructs the Wigner function.}
    \label{fig:principal}
\end{figure*}

Figure~\ref{fig:principal}(a) schematically illustrates the physical setup considered throughout this work. We focus on a circuit-QED architecture composed of a single bosonic mode of a coplanar-waveguide resonator, with frequency $\omega_c$, dispersively coupled to an $M$-qubit superconducting register. The qubits are assumed to be individually addressable and may be implemented, for instance, as SQUID-based transmons operated as effective two-level systems. In this configuration, each qubit frequency $\omega_{q_k}(\Phi_k)$ can be tuned by a local flux bias $\Phi_k$, allowing the qubit--resonator detuning $\Delta_k(\Phi_k)=\omega_{q_k}(\Phi_k)-\omega_c$ to be calibrated~\cite{Koch2007, Krantz2019, Kjaergaard2020, BlaisRMP2021}. 
    
After preparing the resonator in a desired initial state, the subsequent coherent dynamics are governed by the effective dispersive qubit-resonator interaction. Under the rotating-wave approximation, valid when the qubit and resonator frequencies largely exceed the coupling strengths so that the counter-rotating terms oscillate rapidly and average to zero over the relevant timescales, the dynamics is governed by the multi-qubit Jaynes-Cummings Hamiltonian~\cite{JaynesCummings1963,haroche2006exploring,BlaisRMP2021},
\begin{align}  
\hat H
=
\omega_c\,\hat a^\dagger\hat a
+
\sum_{k=1}^{M} \left[
\frac{\omega_{q_k}(\Phi_k)}{2}\,\hat\sigma_z^{(k)}  +
g_k\hat a\,\hat\sigma_+^{(k)} 
 \right] + \text{h.c.},
\label{eq:HJC_multi}
\end{align}
where $\hat a$ ($\hat a^\dagger$) annihilates (creates) an excitation in the resonator mode, $\omega_{q_k}(\Phi_k)$ is the flux-dependent transition frequency of the $k$-th qubit, and $\hat\sigma_-^{(k)}$ ($\hat\sigma_+^{(k)}$) is the lowering (raising) operator for the corresponding effective two-level system, and $g_k$ characterizes the qubit-resonator interaction strengths.

In what follows, our interest lies in the dispersive regime, in which the qubit and resonator are far detuned,
\begin{equation}
|\Delta_k(\Phi_k)|
\equiv
|\omega_{q_k}(\Phi_k)-\omega_c|
\gg
g_k\sqrt{\bar n+1},
\label{eq:dispersive_condition}
\end{equation}
with $\bar n=\langle \hat n\rangle$ and $\hat n=\hat a^\dagger\hat a$. In this regime, real exchange of excitations between qubit and field is strongly suppressed, while virtual processes induce photon-number-dependent energy shifts~\cite{Schuster2007, BlaisRMP2021}. These shifts are precisely the resource exploited in our protocol, since they allow the bosonic field to imprint number-dependent phases on the qubit register without being directly absorbed.

As shown in Appendix~\ref{ap:effective_hamiltonian}, the dispersive limit of Eq.~\eqref{eq:HJC_multi} can be obtained through a Schrieffer--Wolff (small-rotation) transformation, followed by a truncation of the Baker--Campbell--Hausdorff expansion to second order in $g_k/\Delta_k$~\cite{Schuster2007,BlaisRMP2021}. In the interaction picture, this yields the effective Hamiltonian
\begin{equation}
\hat H_{\rm int}
=
\sum_{k=1}^{M}
\chi_k\,\hat n\,\hat\sigma_z^{(k)},
\label{eq:Hint_main}
\end{equation}
with dispersive shifts $\chi_k(\Phi_k)\simeq |g_k|^2/\Delta_k(\Phi_k)$. Equation~\eqref{eq:Hint_main} is the effective interaction underlying the protocol: during a controlled interaction window, each qubit acquires a phase conditioned on the cavity photon number, while the field energy is preserved. In the circuit implementation considered in this work, the bare couplings $g_k$ are fixed by the device geometry, so the relevant control parameter for the dispersive dynamics is the tunable detuning $\Delta_k(\Phi_k)$, set by the flux-dependent qubit frequency. This flux tunability therefore provides a direct way to calibrate the set of dispersive shifts $\{\chi_k\}$ required for the multi-qubit QPE sequence. More general architectures may instead employ tunable couplers or parametrically controlled interactions to adjust the effective qubit--resonator coupling, and hence the corresponding dispersive shifts, dynamically~\cite{gambetta2011,noh2023}.

Although our main target platform is circuit-QED, where long-lived microwave resonators, high-fidelity superconducting qubits, and controllable dispersive interactions are experimentally available~\cite{Schuster2007,BlaisRMP2021,ripoll2022}, the effective model in Eq.~\eqref{eq:Hint_main} is not restricted to superconducting hardware. The essential requirements are: i) a bosonic mode dispersively coupled to individually addressable effective two-level systems, and ii) sufficient control over the accumulated dispersive phases $\chi_k t_k$ to satisfy the QPE calibration condition. How this phase control is realized, i.e., through tunable detunings, tunable couplers, or qubit-dependent interaction times, is platform dependent. Hence, whenever these two requirements are satisfied, the same formalism can in principle be adapted to other architectures, such as atomic cavity-QED or trapped-ion systems, provided that the required ancilla control and readout operations are available.

\section{Protocol}\label{sec:protocol}

The central ingredient of the present protocol is the Quantum Phase Estimation algorithm~\cite{kitaev_1995}, whose circuit structure is summarized in Fig.~\ref{fig:principal}(b). Its purpose is to estimate the phase $\theta\in[0,1)$ associated with an eigenstate $\ket{u}$ of a unitary operator $\hat U$, defined by
\begin{equation}
\hat U \ket{u}=e^{2\pi i\theta}\ket{u}.
\label{eq:qpe_problem}
\end{equation}
In the standard formulation of QPE, an $M$-qubit register is first prepared in a uniform superposition by Hadamard gates, then the $k$-th qubit controls the power $\hat U^{2^{k-1}}$, and a final Fourier-decoding stage maps the accumulated relative phases onto a computational basis outcome that approximates $2^M\theta$. In our case, this general logic is adapted to estimate photon-number dependent phases imprinted on a multi-qubit register by the dispersive interaction with the resonator field.

To describe the ideal action of the protocol on a photon-number eigenstate, we consider the field initially prepared in a Fock state $\ket{n}$, while the qubits are prepared in the ground state $\ket{0}^{\otimes M}$. After applying Hadamard gates to all $M$ qubits, the register is prepared in the uniform superposition
\begin{equation}
\ket{\psi_R}
=
\frac{1}{\sqrt{2^M}}
\sum_{x=0}^{2^M-1}\ket{x},
\label{eq:uniform_superposition}
\end{equation}
where $\ket{x}\equiv\ket{b_{M-1}\cdots b_0}$ denotes the computational basis of the register, with $x=\sum_{j=0}^{M-1}b_j2^j$ and $b_j\in\{0,1\}$. The joint state after this step is $\ket{\psi_R}\otimes\ket{n}$. This superposition allows the subsequent controlled evolution to encode the eigenphase coherently across the register amplitudes.

To tailor QPE to the photon-number problem, it is convenient to define the unitary
\begin{equation}
\hat U_{\rm QPE}=\exp\!\left(\frac{-2\pi i\hat n}{N_{\max}}\right),
\label{eq:U_qpe}
\end{equation}
where $\hat n=\hat a^\dagger \hat a$ is the number operator of the bosonic mode and $N_{\rm max}=2^M$. Since the Fock states are eigenstates of $\hat n$, they are also eigenstates of $\hat U_{\rm QPE}$:
\begin{equation}
\hat U_{\rm QPE}\ket{n}=e^{-2\pi i n/N_{\max}}\ket{n}.
\label{eq:uqpe_fock_eigenvalue}
\end{equation}
Thus, the resonator photon number is encoded as a phase on the unit circle, with resolution set by the size of the register. In this sense, $N_{\max}$ denotes the maximum number of distinguishable phase values that the protocol can resolve.

The QPE block maps photon-number information onto the qubit register through the controlled powers of Eq.~\eqref{eq:U_qpe}, as represented in Fig.~\ref{fig:principal}(b). As shown explicitly in Appendix~\ref{app:uqpe_from_hint}, after the controlled-phase stage each Fock component $\ket{n}$ produces the register state
\begin{equation}
\ket{\psi_R^{(n)}}
=
\frac{1}{\sqrt{N_{\max}}}
\sum_{x=0}^{N_{\max}-1}
e^{-2\pi i nx/N_{\max}}\ket{x}.
\label{eq:phase_gradient_register}
\end{equation}
The inverse quantum Fourier transform~\cite{cleve1998quantum,Torosov2009,nielsen2010quantum} decodes this phase gradient into a computational-basis output $y\in\{0,\dots,N_{\max}-1\}$ correlated with the photon number. For a general field state whose relevant photon-number support lies within $N_{\max}$, the register outcome is therefore unambiguously correlated with the corresponding Fock component, and measuring $y=n$ projects the field onto $\ket{n}$.

The physical implementation of these controlled powers follows directly from the dispersive interaction of Eq.~\eqref{eq:Hint_main}. Up to irrelevant global phases, this interaction can be written in a controlled-phase form acting on the excited branch of each qubit. Then, for a single qubit initially prepared in a superposition $\ket{+}_k=(\ket{0_k}+\ket{1_k})/\sqrt{2}$, and a field Fock component $\ket{n}$, the dispersive evolution during an interaction time $t_{\rm int}$ yields
\begin{equation}
\ket{+}_k\ket{n}\ \longrightarrow\
\frac{\ket{0_k}+e^{-i\chi_k n t_{\rm int}}\ket{1_k}}{\sqrt{2}}\ket{n}.
\label{eq:single_qubit_phase}
\end{equation}
Therefore, each qubit acquires a photon-number dependent relative phase while the field energy is preserved.

To reproduce the logical action of QPE, the physical phase in Eq.~\eqref{eq:single_qubit_phase} must match the phase associated with the controlled power $\hat U_{\rm QPE}^{2^{k-1}}$. Using Eq.~\eqref{eq:uqpe_fock_eigenvalue}, this condition reads
\begin{equation}
\hat U_{\rm QPE}^{2^{k-1}}\ket{n}
=
\exp\!\left(\frac{-2\pi i2^{k-1}n}{N_{\max}}\right)\ket{n},
\end{equation}
which leads to the calibration rule
\begin{equation}
\chi_k\,t_{\rm int}
=
\frac{2\pi\,2^{k-1}}{N_{\max}}.
\label{eq:calibration_main}
\end{equation}
As detailed in Appendix~\ref{app:uqpe_from_hint}, this calibration provides the link between QPE and its circuit-QED realization. Since $\hat U_{\rm QPE}$ is generated by the photon-number operator, the power $2^{k-1}$ is implemented by matching the accumulated dispersive phase rather than by applying $2^{k-1}$ sequential gates. In the flux-tunable realization introduced in Sec.~\ref{sec:model}, we use a common interaction time $t_{\rm int}$ and calibrate the required $\chi_k$ through the qubit-resonator detunings $\Delta_k(\Phi_k)$. More generally, the protocol only requires the accumulated phases $\chi_k t_k$ to satisfy Eq.~\eqref{eq:calibration_main}, irrespective of the specific hardware mechanism used to achieve them.

A key feasibility condition for the protocol is that the dispersive interaction window remains short compared with the cavity lifetime, $t_{\rm int}\ll \kappa^{-1}$, while the dispersive approximation remains valid over all significantly populated photon-number components. For a Fock component $\ket{n}$, this requires $(g_k/\Delta_k)\sqrt{\bar n+1}\ll1$~\cite{Schuster2007, BlaisRMP2021}. This condition is commonly expressed in terms of the critical photon number,
$n_{\rm crit}^{(k)}\simeq \Delta_k^2/(4g_k^2)$, above which higher-order corrections to the dispersive expansion become important~\cite{Schuster2007}. Accordingly, the relevant photon-number support should remain both within the QPE resolved range $n<N_{\max}$ and well below $n_{\rm crit}^{(k)}$ for all qubits participating in the readout.

Representative circuit-QED operating points involve qubit-resonator coupling strengths of a few tens of MHz and detunings of order $1~\mathrm{GHz}$~\cite{BlaisRMP2021, Krantz2019}. As an illustrative choice within this range, we take $g_k/2\pi=30~\mathrm{MHz}$ and $|\Delta_k|/2\pi=2~\mathrm{GHz}$~\cite{BlaisRMP2021}, which give $n_{\rm crit}^{(k)}\simeq1.1\times10^3$ and $|\chi_k|/2\pi\simeq0.45~\mathrm{MHz}$ within the two-level dispersive approximation. The largest controlled phase required by Eq.~\eqref{eq:calibration_main} is $\chi_{\max}t_{\rm int}=\pi$. Thus, for representative dispersive shifts $|\chi_{\max}|/2\pi\sim0.1$--$1~\mathrm{MHz}$, the corresponding interaction window is $t_{\rm int}\sim0.5$--$5~\mu\mathrm{s}$. These times are shorter than typical superconducting-qubit coherence times~\cite{Krantz2019, Kjaergaard2020} and well below the coherence times demonstrated for high-$Q$ superconducting cavities~\cite{Reagor2016, Milul2023}, supporting the separation of timescales assumed in the protocol.

Finally, because the protocol estimates the phase modulo $2\pi$, the measurement outcome resolves the photon number modulo $N_{\max}$. Consequently, $\ket{n}$ and $\ket{n+l N_{\max}}$, with integer $l$, are indistinguishable under the encoding of Eq.~\eqref{eq:U_qpe}. In practice, this ambiguity is suppressed by choosing $N_{\max}$ large enough to cover the relevant photon-number support. Under this condition, measuring the QPE register realizes a number-resolved QND measurement: the field is projected onto the photon-number component associated with the measured register outcome without requiring direct photon absorption.

\section{QND monitoring}
\label{sec:qnd_monitoring_results}

We now use the QPE block as a QND monitor of the dissipative dynamics of a resonator field, following the sequence illustrated in Fig.~\ref{fig:principal}(c). The goal is to track the photon-number evolution of an initially prepared Fock state while the resonator undergoes photon loss. Each QPE interrogation is performed during a short dispersive interaction window $t_{\rm int}\ll\kappa^{-1}$ and is followed by a longer free-evolution interval $\Delta t\gg t_{\rm int}$, during which the cavity field evolves under its dissipative dynamics before the next measurement. This setting is directly inspired by the cavity-QED experiments of Ref.~\cite{Guerlin2007}, where repeated QND measurements revealed both the progressive collapse of the field and the subsequent staircase-like decay of the photon number through quantum jumps. In the present protocol, the atomic Ramsey probes are replaced by a multi-qubit QPE register: photon-number information is encoded into the register through the dispersive interaction, while the resonator is repeatedly interrogated without direct photon absorption.

Each monitoring cycle consists of two stages. First, the QPE interaction is applied during a short time window $t_{\rm int}$, chosen such that $t_{\rm int}\ll \kappa^{-1}$, where $\kappa$ is the field decay rate. Second, after the QPE interrogation, the system evolves freely for a longer interval $\Delta t$, during which photon loss can occur. In the simulations shown below, we define a total monitoring time $T_{\rm total}=t_{\rm cyc}N_{\rm cyc}$, where $N_{\rm cyc}$ is the number of QPE cycles and $t_{\rm cyc}=T_{\rm total}/N_{\rm cyc}$ the duration of one monitoring cycle. We take $T_{\rm total}=5\kappa^{-1}$ and $N_{\rm cyc}=100$, and split each cycle as $t_{\rm cyc}=t_{\rm int}+\Delta t$. The QPE interaction occupies only $2\%$ ($f_{\rm int}=1/50$) of each cycle. Thus, $ t_{\rm int}=f_{\rm int}t_{\rm cyc}$, where this choice makes the QPE interrogation much shorter than the free-evolution interval.

\begin{figure}[htb!]
    \centering
    \includegraphics[width=1.0\linewidth]{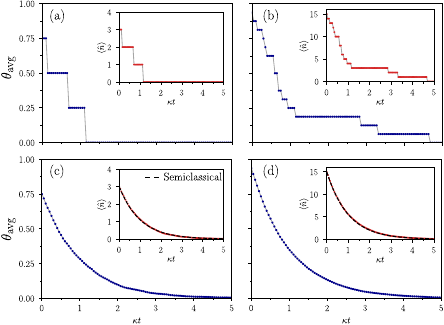}
    \caption{\justifying QND monitoring of the dissipative field dynamics using QPE. The main panels show the average phase
    $\theta_{\rm avg}(t)=\sum_y \theta_y P_y(t)$, with $\theta_y=y/2^M$, obtained from the QPE register after each free-evolution step of the cavity field. Panels (a) and (b) display single quantum trajectories, while panels (c) and (d) show averages over $1000$ trajectories. In (a) and (c), the field is initialized in the Fock state $\ket{3}$ and monitored with an $M=2$ qubits. In (b) and (d), the field is initialized in $\ket{15}$ and monitored with an $M=4$ qubits. The insets show the corresponding mean photon number, $\langle \hat n\rangle=\langle \hat a^\dagger \hat a\rangle$, measured during the same monitoring process. For the averaged trajectories in (c) and (d), the dashed black curves in the insets represent the semiclassical intensity benchmark $|\alpha_{\rm sc}(t)|^2=N e^{-\kappa t}$, where $N$ is the initial photon number.}
    \label{fig:qpe_monitoring}
\end{figure}

The numerical results were obtained from a quantum-trajectory simulation of the full open-system dynamics. The cavity loss channel is included through the collapse operator $\hat C=\sqrt{\kappa}\hat a$, and the calibrated dispersive Hamiltonian is switched on only during the QPE interaction window $t_{\rm int}$. In this sense, the simulation does not implement an abstract gate-level circuit alone. Instead, it follows the continuous-time dissipative evolution that realizes the controlled QPE phases through the physical interaction. After each QPE window, the inverse quantum Fourier transform is applied to the qubit register and the probabilities $P_y(t)$ are extracted by projection onto the computational basis. The monitored quantity plotted in the main panels of Fig.~\ref{fig:qpe_monitoring} is the average phase extracted from the QPE register,
\begin{equation}
    \theta_{\rm avg}(t)
    =
    \sum_{y=0}^{2^M-1}
    \theta_y P_y(t),
    \label{eq:theta_avg_monitoring}
\end{equation}
where $\theta_y=y/2^M$ and $P_y(t)$ is the probability of obtaining the register outcome $y$ after the inverse quantum Fourier transform. For a Fock component $\ket{n}$ whose photon number lies inside the resolved interval $0\le n<N_{\max}=2^M$, the QPE distribution is concentrated around $y=n$. Therefore, in the ideal resolved limit,
\begin{equation}
    \theta_{\rm avg}(t) \simeq \frac{\langle \hat n(t)\rangle}{N_{\max}}.
    \label{eq:theta_n_relation}
\end{equation}
This relation makes $\theta_{\rm avg}$ a direct normalized estimator of the photon number.

\begin{figure*}[htb!]
    \centering
    \includegraphics[width=1.0\linewidth]{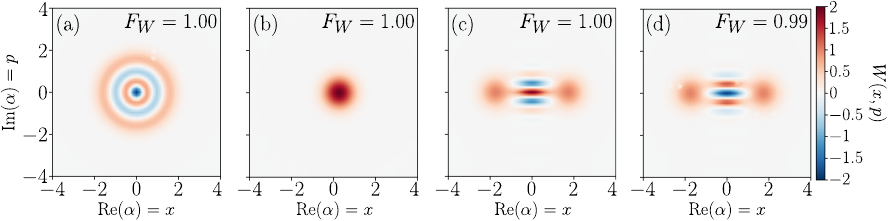}
    \caption{\justifying Wigner function reconstruction in the phase-space diagram from the displaced photon-number distributions obtained by repeating the QPE readout. We use an $M=6$ qubits, corresponding to $N_{\max}=2^M=64$, and consider $N=3$ in all panels: (a) a Fock state $\ket{N}$, (b) a coherent state $\ket{\alpha_0}$ with $\alpha_0=\sqrt{N}$, and (c,d) even and odd Schrödinger-cat states, $|\mathcal C_\pm(\alpha_0)\rangle=\mathcal N_\pm(|\alpha_0\rangle\pm|-\alpha_0\rangle)$, respectively. The horizontal and vertical axes denote the real and imaginary parts of the displacement amplitude, $\alpha=x+ip$. The inset value in each panel is the Wigner-overlap fidelity $\mathcal F_W$.}
    \label{fig:qpe_tomography}
\end{figure*}

Figures~\ref{fig:qpe_monitoring}(a) and \ref{fig:qpe_monitoring}(b) show single monitored quantum trajectories. In Fig.~\ref{fig:qpe_monitoring}(a), the field is initially prepared in $\ket{3}$ and monitored with an $M=2$ qubit register, so that $N_{\max}=4$. The initial phase is therefore close to $\theta_{\rm avg}=3/4$, and each photon loss changes the phase by one register unit, $\Delta\theta=1/N_{\max}=1/4$. The trajectory displays the expected staircase pattern: $\ket{3}\rightarrow \ket{2}\rightarrow \ket{1}\rightarrow \ket{0}$, with the field eventually reaching the vacuum state. The inset shows the corresponding evolution of the mean photon number, $\langle \hat n\rangle=\langle \hat a^\dagger\hat a\rangle$, which exhibits the same discrete jumps.

Figure~\ref{fig:qpe_monitoring}(b) illustrates the same monitoring process for a larger initial state, $\ket{15}$, using an $M=4$ qubit register, so that $N_{\max}=16$. In this case the register resolution is $\Delta\theta=1/16$, and the initial QPE phase is $\theta_{\rm avg}=15/16$. Since the photon-loss rate out of a Fock state $\ket{n}$ is proportional to $n\kappa$, the first jumps are more likely to occur rapidly, while the lower-photon-number states tend to survive for longer times. This behavior is the QPE counterpart of the staircase-like decay observed in the photon-counting of Ref.~\cite{Guerlin2007}. The essential feature is the discreteness of the loss channel: in a quantum trajectory, the collapse operator $\hat a$ maps $\ket{n}$ onto $\ket{n-1}$, so each photon-loss event shifts the QPE output by one register bin. The plateau lengths are stochastic and depend on the instantaneous photon number, since the jump rate out of $\ket{n}$ is proportional to $n\kappa$.

The averaged behavior is shown in Figs.~\ref{fig:qpe_monitoring}(c) and \ref{fig:qpe_monitoring}(d), where the same observables are averaged over $1000$ quantum trajectories. The individual stochastic jumps are washed out by ensemble averaging, yielding a smooth decay of both $\theta_{\rm avg}(t)$ and $\langle \hat n(t)\rangle$. The dashed black curves in the insets correspond to the semiclassical intensity benchmark
\begin{equation}
    |\alpha_{\rm sc}(t)|^2=N e^{-\kappa t},
    \label{eq:semiclassical_decay_main}
\end{equation}
where $N$ is the initial photon number. A derivation of Eq.~\eqref{eq:semiclassical_decay_main} from the Lindblad dynamics is given in Appendix~\ref{ap:semiclassical_decay}. The agreement between the averaged quantum trajectories and this exponential law confirms that the QPE measurement itself does not introduce spurious energy relaxation: the decay is governed by the physical cavity-loss channel. Equivalently, using Eq.~\eqref{eq:theta_n_relation}, the averaged phase follows
\begin{equation}
    \theta_{\rm avg}(t)
    \simeq
    \frac{N}{N_{\max}}e^{-\kappa t}.
    \label{eq:theta_semiclassical_decay}
\end{equation}

These results demonstrate that QPE can be used as a QND method for monitoring photon-number dynamics. In a single trajectory, the protocol resolves individual loss events as discrete jumps in the inferred photon number. In an ensemble average, the same dynamics reproduce the expected exponential relaxation of the field energy. Thus, the QPE readout provides photon-number information without directly absorbing the field, while the irreversible dynamics is governed by the physical cavity-loss channel.

\section{Wigner Tomography}
\label{sec:tomography}

The monitoring results discussed in the previous section show that the QPE block can be used as a nondestructive photon-number readout. We now use the same QPE block for a different task: reconstructing the Wigner function of the resonator field, following the displaced-QPE sequence illustrated in Fig.~\ref{fig:principal}(d). The central observation is that the Wigner function at a given phase-space point $\alpha$ can be obtained from the photon-number parity of the corresponding displaced field state~\cite{lutterbach1997direct, deleglise2008reconstruction}. Therefore, by first applying a controlled displacement $\hat D(\alpha)$ and then repeating the QPE readout, one can estimate the displaced photon-number distribution $P_n(\alpha)$ and evaluate its parity. Repeating this procedure for different values of $\alpha$ reconstructs the Wigner function over phase space.

For each phase-space point $\alpha=x+ip$, the tomography sequence is as follows. First, the resonator is prepared in the target state $\rho$. Then, a displacement operation $D(-\alpha)$ is applied to the field. The multi-qubit register is initialized, the calibrated dispersive QPE block is performed, and the register is measured after the inverse quantum Fourier transform. A single run produces one register outcome $y$, which is interpreted as a photon-number sample of the displaced state, up to the finite resolution set by $N_{\max}=2^M$. Repeating the same preparation--displacement--QPE sequence many times for the same value of $\alpha$ yields the displaced photon-number distribution $P_y^{\rm QPE}(\alpha)$. The Wigner function is then reconstructed from the alternating parity sum
\begin{equation}
W_{\rm QPE}(\alpha)
=
2\sum_{y=0}^{N_{\max}-1}
(-1)^y P_y^{\rm QPE}(\alpha),
\label{eq:W_QPE_main}
\end{equation}
where we use the convention in which $W(\alpha)$ ranges from $-2$ to $2$. A detailed derivation of Eq.~\eqref{eq:W_QPE_main}, including the displacement-sign convention and the connection with the parity operator, is given in Appendix~\ref{ap:wigner_qpe}.

This tomographic protocol differs from the monitoring sequence in an important way. In the monitoring task, each QPE interrogation is followed by a free-evolution interval $\Delta t$, during which the field can lose photons and the next QPE cycle probes the updated state. In tomography, by contrast, no intentional free-evolution interval is inserted between the displacement and the QPE readout. Each shot samples the photon-number distribution of the same displaced target state. Thus, the repetitions required to estimate $P_y^{\rm QPE}(\alpha)$ should be understood as repeated preparations of the same state, or equivalently as independent experimental realizations with the same displacement. This distinction is important because a projective QPE measurement, although nondestructive with respect to photon absorption, still collapses the displaced field onto a photon-number component.

The statistical cost of one phase-space point can be estimated directly from Eq.~\eqref{eq:W_QPE_main}. If $N_{\rm shots}$ independent QPE shots are performed at a fixed displacement $\alpha$, an unbiased parity estimator is
\begin{equation}
\widehat W(\alpha)
=
\frac{2}{N_{\rm shots}}
\sum_{r=1}^{N_{\rm shots}}
(-1)^{y_r},
\label{eq:W_estimator}
\end{equation}
where $y_r$ is the register outcome in the $r$-th repetition. As shown in Appendix~\ref{ap:wigner_qpe}, its variance is
\begin{equation}
{\rm Var}\!\left[\widehat W(\alpha)\right]
=
\frac{4-W^2(\alpha)}{N_{\rm shots}}
\leq
\frac{4}{N_{\rm shots}}.
\label{eq:W_variance}
\end{equation}
Here we define $\varepsilon_W$ as the target standard deviation statistical uncertainty of the estimator, i.e.,
${\rm Var}[\widehat W(\alpha)]\leq \varepsilon_W^2$. Since the Wigner function satisfies $-2\leq W(\alpha)\leq 2$, the variance is maximal at phase-space points where $W(\alpha)=0$. A conservative worst-case requirement is therefore $N_{\rm shots}\gtrsim 4/\varepsilon_W^2$. For example, $\varepsilon_W\simeq 0.05$ requires on the order of $10^3$ QPE shots per phase-space point, while $\varepsilon_W\simeq 0.02$ requires on the order of $10^4$ shots. In regions where $|W(\alpha)|$ is close to its extremal value $2$, fewer repetitions are needed because the parity outcomes become less fluctuating. Also, this estimate refers only to projection noise in the parity sampling. Additional experimental errors, such as imperfect displacements, qubit readout errors, decoherence during the QPE window, and calibration errors in $\chi_k t_{\rm int}$, would contribute extra uncertainty.

Another important resource is the number of qubits in the QPE register. The register size fixes $N_{\max}=2^M$, which determines the number of distinct photon-number outcomes resolved by the algorithm. Since the protocol is applied after phase-space displacements, the relevant photon-number support is not only that of the original state $\rho$, but that of the displaced state $D(-\alpha)\rho D(\alpha)$ for every sampled $\alpha$. If this displaced distribution has significant weight outside the Hilbert-space window used in the reconstruction, the estimated distribution and the Wigner function may be distorted. In the present implementation, we choose a number of qubits that produces an $N_{\max}$ large enough to contain the relevant photon-number support over the whole phase-space grid. 

Figure~\ref{fig:qpe_tomography} shows the resulting QPE-based Wigner reconstructions for representative bosonic states with $N=3$. Panel~\ref{fig:qpe_tomography}(a) corresponds to a Fock state $\ket{N}$, whose Wigner function displays the expected ring-like structure and alternating positive and negative regions, revealing its strongly nonclassical character. Panel~\ref{fig:qpe_tomography}(b) shows the coherent state $\ket{\alpha_0}$ with $\alpha_0=\sqrt{N}$, which produces a positive, Gaussian-like Wigner distribution in phase space. Panels~\ref{fig:qpe_tomography}(c) and~\ref{fig:qpe_tomography}(d) show even and odd Schrödinger-cat states,
\begin{equation}
|\mathcal C_\pm(\alpha_0)\rangle
=
\mathcal N_\pm
\left(
|\alpha_0\rangle
\pm
|-\alpha_0\rangle
\right),
\end{equation}
respectively. In these cases, the two positive lobes represent the quasi-classical coherent components, while the oscillatory fringes between them encode their quantum coherence. The sign change of the central interference pattern distinguishes the even and odd cats.

To quantify the reconstruction quality, the inset value in each panel reports the Wigner-overlap fidelity $\mathcal F_W$, defined in Appendix~\ref{ap:wigner_qpe}. This quantity compares the QPE-reconstructed Wigner function with an independently computed ideal Wigner function on the same phase-space grid. In the present numerical setting, $\mathcal F_W$ should be interpreted as a benchmark of tomographic accuracy under finite grid size, finite Hilbert-space cutoff, and numerical solver tolerances, rather than as an independent experimental certification of state preparation. The near-unity values shown in Fig.~\ref{fig:qpe_tomography} indicate that, for the chosen register size and phase-space window, the QPE readout accurately reproduces the displaced parity structure of all four target states.

It is useful to place this result in the context of cavity-QED tomography. In the experiments of Refs.~\cite{Guerlin2007,  deleglise2008reconstruction}, photon-number information was extracted from long sequences of Rydberg atoms crossing a high-$Q$ microwave cavity. In particular, QND photon counting enabled the observation of progressive field-state collapse and staircase-like quantum jumps~\cite{Guerlin2007}, while displaced-field measurements enabled the reconstruction of coherent, Fock, and Schrödinger-cat states~\cite{nogues2000measurement, bertet2002direct, deleglise2008reconstruction}. The present protocol follows the same physical logic: displace the field, measure number-sensitive information, and infer the displaced parity. However, it replaces the sequential atomic Ramsey probes with a multi-qubit QPE register. In this sense, the role of QPE is not merely to provide another way of measuring parity, but to implement a number-resolved QND readout: the same register measurement yields the photon-number distribution of the displaced field and can therefore be used for both photon-number monitoring and Wigner tomography. The comparison should not be understood as claiming experimental superiority over those cavity-QED methods, nor over parity-only readout schemes. Rather, QPE offers a unified circuit-based alternative that is naturally matched to superconducting architectures, where multi-qubit registers, dispersive couplings, fast readout, and controlled displacements are well-developed tools.

\section{Conclusion}\label{sec:conclusion}

We have proposed a circuit-QED protocol that uses quantum phase estimation as a unified dispersive readout primitive for QND monitoring and Wigner tomography of a bosonic field. The central idea is to encode photon-number information into the phases accumulated by a multi-qubit register through the dispersive interaction with a resonator mode. The register measurement provides a number-resolved output without requiring direct absorption of the resonator photons. In this way, QPE can be interpreted not only as an algorithmic subroutine, but also as a physically meaningful measurement tool for bosonic quantum systems.

For the monitoring task, we showed that repeated QPE interrogations can track the dissipative photon-number dynamics of an initially prepared Fock state. In single quantum trajectories, photon-loss events appear as discrete jumps in the inferred photon number, reproducing the staircase-like behavior characteristic of QND photon counting. In ensemble averages, the same dynamics recover the expected exponential relaxation of the field energy, confirming that the decay is governed by the physical cavity-loss channel rather than by the QPE readout itself. This result should be viewed as a circuit-QED counterpart and complementary proposal to the seminal cavity-QED experiments in which long sequences of Rydberg atoms were used to monitor progressive field-state collapse and photon-number quantum jumps~\cite{Guerlin2007}. Here, the role of the atomic Ramsey probes is replaced by a calibrated multi-qubit register, making the protocol naturally compatible with superconducting quantum hardware.

We then showed that the same QPE block can be reused for Wigner tomography. By applying controlled phase-space displacements before the QPE readout, the protocol samples the displaced photon-number distribution and reconstructs the Wigner function through the corresponding parity sum. We demonstrated this procedure for representative states, including Fock, coherent, and even/odd Schrödinger cat states. The Wigner-overlap fidelities are close to unity in the numerical implementation.

The proposal is especially relevant for quantum computing with bosonic modes. In circuit-QED, Wigner tomography and nondestructive monitoring are central tools for calibrating and controlling bosonic memories~\cite{Heeres2015, Eickbusch2022}, verifying nonclassical resource states~\cite{bertet2002direct, deleglise2008reconstruction, vlastakis2013cat}, diagnosing photon loss and quantum jumps~\cite{Guerlin2007, sun2014tracking}, and validating bosonic error-correction or state-preparation protocols~\cite{Michael2016, Ofek2016, CampagneIbarcq2020}. The present results show that the same dispersive QPE architecture can perform both tasks: monitoring photon-number trajectories through repeated nondestructive readout and reconstructing phase-space structure through displaced QPE measurements. In this sense, the protocol provides a unified route to QND monitoring and tomography in a hardware platform directly connected to bosonic quantum information processing.

Finally, the same QPE-based framework naturally extends from measurement to state engineering. In Ref.~\cite{lucas2026largefockstates}, dispersive photon-number encoding through QPE is combined with quantum amplitude amplification to generate large Fock states with high probability. Together with the monitoring and tomography protocols developed here, this result points toward a broader framework in which QPE-based dispersive interactions can be used to prepare, monitor, and characterize nonclassical bosonic states within closely related circuit-QED architectures.

\section{Acknowledgments}

The authors thank Luiz Otavio Ribeiro Solak, Alan Costa dos Santos and Juan José García-Ripoll for valuable discussions and suggestions on this work. This study was financed, in part, by the São Paulo Research Foundation (FAPESP), Brazil, Process Number \#2022/00209-6, \#2025/15490-0, \#2024/02604-5, \#2022/10218-2 and \#2025/23694-5, by the Coordenação de Aperfeiçoamento de Pessoal de Nível Superior (CAPES) Finance Code 001, and by the Brazilian National Council for Scientific and Technological Development -- CNPq, Grants No.~140001/2023-9, No.~405712/2023-5, No.~311612/2021-0, and No. 302234/2026-8.

\appendix

\section{Effective dispersive Hamiltonian}\label{ap:effective_hamiltonian}

To derive the effective interaction used throughout this work, it is convenient to first consider a single qubit. The corresponding Jaynes--Cummings Hamiltonian of Eq.~\eqref{eq:HJC_multi} can be written as follows:
\begin{equation}
\hat H_{\rm JC}
=
\omega_c\,\hat a^\dagger\hat a
+
\frac{\omega_q}{2}\,\hat\sigma_z
+
g\left(
\hat a\,\hat\sigma_+
+
\hat a^\dagger\hat\sigma_-
\right).
\label{eq:HJC_single}
\end{equation}
We separate it into a free part and an interaction part as
\begin{align}
\hat H_{\rm JC} &= \hat H_0 + \hat V,\\
\hat H_0 &= \omega_c\,\hat a^\dagger\hat a + \frac{\omega_q}{2}\,\hat\sigma_z,\\
\hat V &= g\left(\hat a\,\hat\sigma_+ + \hat a^\dagger\hat\sigma_-\right).
\end{align}
In the dispersive regime, the natural small parameter is $g/\Delta$, where $\Delta=\omega_q-\omega_c$. One may then eliminate the exchange term perturbatively through a Schrieffer--Wolff (or small-rotation) transformation~\cite{SchriefferWolff1966, Brion2007, Schuster2007}, as $\hat U = e^{\hat S}$ with
\begin{equation}
\hat S
=
\frac{g}{\Delta}
\left(
\hat a\,\hat\sigma_+
-
\hat a^\dagger\hat\sigma_-
\right),
\label{eq:SW_generator_single}
\end{equation}
where $\hat S^\dagger=-\hat S$, ensuring the unitarity of $\hat U$.

The transformed Hamiltonian is obtained from the Baker--Campbell--Hausdorff expansion,
\begin{equation}
\hat H_{\rm eff}
=
\hat U\hat H_{\rm JC}\hat U^\dagger
=
\hat H_{\rm JC}
+
[\hat S,\hat H_{\rm JC}]
+
\frac{1}{2}[\hat S,[\hat S,\hat H_{\rm JC}]]
+\cdots.
\label{eq:BCH_expansion}
\end{equation}
The generator $\hat S$ is chosen such that
\begin{equation}
[\hat S,\hat H_0] = -\hat V,
\end{equation}
so that the exchange interaction is cancelled at first order in $g/\Delta$. Truncating the expansion at second order because of the dispersive condition $g/\Delta \ll 1$, one obtains
\begin{equation}
\hat H_{\rm eff}
\simeq
\hat H_0
+
\frac{1}{2}[\hat S,\hat V].
\end{equation}
Evaluating the commutator yields the standard dispersive Hamiltonian~\cite{Schuster2007,BlaisRMP2021}
\begin{equation}
\hat H_{\rm eff}
=
\omega_c\,\hat a^\dagger\hat a
+
\frac{1}{2}\left(\omega_q+\chi\right)\hat\sigma_z
+
\chi\,\hat a^\dagger\hat a\,\hat\sigma_z
+
\frac{\chi}{2}\,I,
\label{eq:Heff_single_full}
\end{equation}
where $\chi=g^2/\Delta$. The last term in Eq.~\eqref{eq:Heff_single_full} is a global energy shift and may be discarded. The second term corresponds to the Lamb/Stark-renormalized qubit frequency, while the third term is the desired dispersive coupling: it shifts the qubit transition frequency proportionally to the intracavity photon number, and equivalently shifts the resonator frequency depending on the qubit state \cite{BlaisRMP2021}.

The generalization to the $M$-qubit case follows directly. Starting from Eq.~\eqref{eq:HJC_multi}, one applies the multi-qubit generator
\begin{equation}
\hat S
=
\sum_{k=1}^{M}
\frac{g_k}{\Delta_k}
\left(
\hat a\,\hat\sigma_+^{(k)}
-
\hat a^\dagger\hat\sigma_-^{(k)}
\right),
\label{eq:SW_generator_multi}
\end{equation}
with $\Delta_k=\omega_{q_k}-\omega_c$, and again expands the transformed Hamiltonian up to second order in $g_k/\Delta_k$. Neglecting higher-order terms and absorbing global energy shifts into the zero of energy, one obtains
\begin{align}
\hat H_{\rm eff}
&=
\omega_c\,\hat a^\dagger\hat a
+
\sum_{k=1}^{M}
\frac{\tilde\omega_{q_k}}{2}\,\hat\sigma_z^{(k)}
+
\sum_{k=1}^{M}
\chi_k\,\hat a^\dagger\hat a\,\hat\sigma_z^{(k)}
+
\hat H_{\rm qq},
\label{eq:Heff_multi_full}
\end{align}
where $\chi_k \simeq g_k^2/\Delta_k$ and $\tilde\omega_{q_k}$ denotes the renormalized qubit frequency. The term $\hat H_{\rm qq}$ collects cavity-mediated qubit--qubit couplings that may appear at the same perturbative order. In the protocol considered here, these additional terms can be neglected when the qubits are individually addressed and effectively coupled one at a time to the resonator, or when residual qubit--qubit interactions are sufficiently small compared to the dispersive shifts of interest. Under these conditions, the relevant entangling dynamics are governed by the photon-number-dependent phase accumulation encoded in the term proportional to $\chi_k$.

Passing now to the interaction picture with respect to the renormalized free Hamiltonian,
\begin{equation}
\hat H_{\rm eff,0}
=
\omega_c\,\hat a^\dagger\hat a
+
\sum_{k=1}^{M}
\frac{\tilde\omega_{q_k}}{2}\,\hat\sigma_z^{(k)},
\end{equation}
the remaining nontrivial interaction is simply
\begin{equation}
\hat H_{\rm int}
=
\sum_{k=1}^{M}
\chi_k\,\hat n\,\hat\sigma_z^{(k)},
\label{eq:Hint_main_ap}
\end{equation}
where $\hat n=\hat a^\dagger\hat a$. Equation~\eqref{eq:Hint_main} is the effective Hamiltonian underlying the whole protocol developed in this work. It shows that the bosonic mode does not exchange excitations with the qubit register. Instead, it imprints photon-number-dependent phases on the qubits. This is precisely the mechanism required for quantum phase estimation and for the nondestructive extraction of information about the field.

\section{Quantum phase estimation and calibration from the dispersive interaction}
\label{app:uqpe_from_hint}

In this Appendix, we present the algebraic derivation of the QPE algorithm used in the main text, as well as the calibration condition that connects the logical controlled powers of $\hat U_{\rm QPE}$ to the physical dispersive interaction.

\subsection{QPE as a Fourier decoding of photon-number phases}

The phase estimation problem consists in determining the phase $\theta\in[0,1)$ associated with an eigenstate $\ket{u}$ of a unitary operator $\hat U$,
\begin{equation}
\hat U\ket{u}=e^{2\pi i\theta}\ket{u}.
\end{equation}
In our case, the relevant unitary is
\begin{equation}
\hat U_{\rm QPE}=\exp\!\left(-\frac{2\pi i\hat n}{N_{\max}}\right),
\end{equation}
whose eigenstates are the Fock states,
\begin{equation}
\hat U_{\rm QPE}\ket{n}=e^{-2\pi i n/N_{\max}}\ket{n}.
\end{equation}
Thus, for a given Fock component $\ket{n}$ the associated phase is
\begin{equation}
\theta_n=\frac{n}{N_{\max}}.
\end{equation}

We now derive the QPE circuit step by step. Let the register contain $M$ qubits, with $N_{\max}=2^M$. The initial state for a Fock component $\ket{n}$ of the field is
\begin{equation}
\ket{\Psi_0}=\ket{0}^{\otimes M}\otimes\ket{n}.
\label{eq:app_qpe_state0}
\end{equation}
After applying Hadamard gates to the register, we obtain
\begin{equation}
\ket{\Psi_1}
=
\left(
\frac{1}{\sqrt{N_{\max}}}
\sum_{x=0}^{N_{\max}-1}\ket{x}
\right)\otimes\ket{n}.
\label{eq:app_qpe_state1}
\end{equation}

The next step is the controlled application of the powers
$\hat U_{\rm QPE}^{2^{k-1}}$ for $k=1,\dots,M$. Writing the binary expansion of the register state as
\begin{equation}
x=\sum_{k=1}^{M} b_{k-1}\,2^{k-1},
\end{equation}
where $b_{k-1}\in\{0,1\}$. The total controlled operation applies $\hat U_{\rm QPE}^{x}$ to the field whenever the ancilla register is in $\ket{x}$. Therefore,
\begin{equation}
\ket{\Psi_2}
=
\frac{1}{\sqrt{N_{\max}}}
\sum_{x=0}^{N_{\max}-1}
\ket{x}\otimes \hat U_{\rm QPE}^{x}\ket{n}.
\label{eq:app_qpe_state2}
\end{equation}
Using the eigenvalue equation of $\hat U_{\rm QPE}$, this becomes
\begin{align}
\ket{\Psi_2}
&=
\frac{1}{\sqrt{N_{\max}}}
\sum_{x=0}^{N_{\max}-1}
e^{-2\pi i nx/N_{\max}}
\ket{x}\otimes\ket{n}.
\label{eq:app_qpe_state2_explicit}
\end{align}

At this point the photon number is encoded in the relative phases of the ancilla register. To decode this information, we introduce the following Fourier-transform convention:
\begin{equation}
\mathrm{QFT}_{N_{\max}}\ket{y}
=
\frac{1}{\sqrt{N_{\max}}}
\sum_{x=0}^{N_{\max}-1}
e^{-2\pi i xy/N_{\max}}
\ket{x},
\label{eq:app_qft_def}
\end{equation}
so that
\begin{equation}
\mathrm{QFT}_{N_{\max}}^\dagger\ket{x}
=
\frac{1}{\sqrt{N_{\max}}}
\sum_{y=0}^{N_{\max}-1}
e^{2\pi i xy/N_{\max}}
\ket{y}.
\label{eq:app_iqft_def}
\end{equation}
Applying $\mathrm{QFT}_{N_{\max}}^\dagger$ to the register gives
\begin{equation}
\ket{\Psi_3}
= \frac{1}{\sqrt{N_{\max}}}
\sum_{y=0}^{N_{\max}-1}
e^{2\pi i xy/N_{\max}}
\ket{y}\otimes\ket{n}.
\label{eq:app_qpe_state3}
\end{equation}
Therefore, for an exact eigenphase representable on the $M$-qubit grid, the QPE readout returns the integer
\begin{equation}
y=n\;(\mathrm{mod}\;N_{\max}).
\end{equation}

The previous derivation applies to each Fock component individually. For a general field state
\begin{equation}
\ket{\psi_f}=\sum_{n=0}^{\infty} c_n \ket{n},
\end{equation}
the same sequence yields, by linearity,
\begin{equation}
\ket{\Psi_{\rm out}}
=
\sum_{n=0}^{\infty}
c_n\,
\ket{n\;(\mathrm{mod}\;N_{\max})}\otimes\ket{n}.
\label{eq:app_general_qpe_output}
\end{equation}
Thus, measuring the ancilla register and obtaining the outcome $y$ projects the field onto
\begin{equation}
\ket{\psi_f^{(y)}}
\propto
\sum_{\ell=0}^{\infty}
c_{\,y+\ell N_{\max}}\ket{y+\ell N_{\max}}.
\label{eq:app_aliasing_state}
\end{equation}
Equation~\eqref{eq:app_aliasing_state} makes explicit the modular nature of the protocol: the measurement distinguishes photon numbers only modulo $N_{\max}$. When the relevant support of the initial field satisfies $n<N_{\max}$, the aliasing terms are absent and the outcome $y=n$ heralds the pure Fock component $\ket{n}$.

For a coherent-state input field,
\begin{equation}
\ket{\alpha}
=
e^{-|\alpha|^2/2}
\sum_{n=0}^{\infty}
\frac{\alpha^n}{\sqrt{n!}}\ket{n},
\end{equation}
the output of the QPE block is therefore
\begin{equation}
\ket{\Psi_{\rm out}}
=
e^{-|\alpha|^2/2}
\sum_{n=0}^{\infty}
\frac{\alpha^n}{\sqrt{n!}}
\ket{n\;(\mathrm{mod}\;N_{\max})}\otimes\ket{n}.
\end{equation}
If $N_{\max}$ is chosen large enough to cover the relevant Poissonian support, measuring the ancilla outcome $y=N$ projects the field onto the target Fock state $\ket{N}$.

\subsection{Controlled-phase form of the dispersive interaction}

We now connect the logical unitary $\hat U_{\rm QPE}$ to the physical qubit--resonator interaction. In the dispersive regime, the effective interaction-picture Hamiltonian was presented in Eq.~\eqref{eq:Hint_main_ap}. Considering a single qubit $k$ case, the atomic inversion operator could be expressed in terms of the excited atomic state $\ket{1_k}$ and the identity operator $\hat I_k$ as
\begin{equation}
\hat\sigma_z^{(k)} = 2\ket{1_k}\!\bra{1_k}-\hat I_k.
\label{eq:app_sigma_projector}
\end{equation}
The corresponding unitary can be rewritten as
\begin{align}
\exp\!\left(-i\chi_k \hat n \hat\sigma_z^{(k)} t\right)
&=
\exp\!\left[-i\chi_k \hat n \left(2\ket{1_k}\!\bra{1_k}-\hat I_k\right)t\right]
\nonumber\\
&=
\exp\!\left(i\chi_k \hat n t\right) \nonumber \\
&\times \exp\!\left(-i\,2\chi_k \hat n \ket{1_k}\!\bra{1_k}t\right).
\label{eq:app_factorized_unitary}
\end{align}
The first exponential in Eq.~\eqref{eq:app_factorized_unitary} acts only on the field and contributes a global phase within each Fock sector. The second exponential is the controlled-phase part relevant to QPE. Since only the relative phase between the qubit basis states matters for the algorithm, it is convenient to absorb this factor of $2$ into the effective calibration parameter and write, for the algorithmic discussion,
\begin{equation}
\hat H_{\rm int}^{(k)}
=
\chi_k\,\hat n\,\ket{1_k}\!\bra{1_k}.
\label{eq:app_hint_projector}
\end{equation}
With this convention, the interaction acts trivially on $\ket{0_k}$ and imprints a phase on $\ket{1_k}$ only:
\begin{align}
\hat U_{\rm int}^{(k)}(t)\ket{0_k}\ket{n}
&=
\ket{0_k}\ket{n},\\
\hat U_{\rm int}^{(k)}(t)\ket{1_k}\ket{n}
&=
e^{-i\chi_k n t}\ket{1_k}\ket{n}.
\end{align}
Therefore, for a qubit initialized in a superposition
\begin{equation}
\ket{+}_k=\frac{\ket{0_k}+\ket{1_k}}{\sqrt{2}},
\end{equation}
we obtain
\begin{equation}
\hat U_{\rm int}^{(k)}(t)\ket{+}_k\ket{n}
=
\frac{\ket{0_k}+e^{-i\chi_k n t}\ket{1_k}}{\sqrt{2}}\ket{n}.
\label{eq:app_single_qubit_phase}
\end{equation}

\subsection{Calibration of the controlled powers}

The $k$-th qubit in QPE must implement the controlled power $\hat U_{\rm QPE}^{2^{k-1}}$. Acting on a Fock state, its effect is
\begin{equation}
\hat U_{\rm QPE}^{2^{k-1}}\ket{n}
=
\exp\!\left(-\frac{2\pi i2^{k-1} n}{N_{\max}}\right)\ket{n}.
\label{eq:app_uqpe_power}
\end{equation}
On the other hand, the physical dispersive interaction gives the phase in Eq.~\eqref{eq:app_single_qubit_phase}. Matching the physical and algorithmic phases requires
\begin{equation}
\exp\left(-i\chi_k n t_{\rm int}\right)
=
\exp\!\left(-\frac{2\pi i2^{k-1} n}{N_{\max}}\right),
\end{equation}
which immediately yields
\begin{equation}
\chi_k\,t_{\rm int}
=
\frac{2\pi\,2^{k-1}}{N_{\max}}.
\label{eq:app_calibration}
\end{equation}

Equation~\eqref{eq:app_calibration} is the calibration condition used in the protocol. It shows that the binary weighting required by QPE is implemented by choosing the effective dispersive coupling of each qubit such that the phase accumulated during the common interaction time $t_{\rm int}$ scales as $2^{k-1}$. Equivalently, for a fixed coupling $\chi_k=\chi$, one may implement the same logical controlled powers by using qubit-dependent interaction times
\begin{equation}
t_k=
\frac{2\pi\,2^{k-1}}{N_{\max}\chi}.
\label{eq:app_time_calibration}
\end{equation}

In summary, the dispersive qubit--resonator interaction naturally realizes the phase-kickback primitive required by QPE: the photon number labels the eigenphase of $\hat U_{\rm QPE}$, and the calibrated interaction of Eq.~\eqref{eq:app_calibration} converts the abstract powers $\hat U_{\rm QPE}^{2^{k-1}}$ into physically implementable controlled-phase operations.

\section{Semiclassical benchmark for the monitored field decay}
\label{ap:semiclassical_decay}

In this Appendix we derive the semiclassical benchmark used in the insets of Fig.~\ref{fig:qpe_monitoring}. The monitored field is subject to photon loss, described by the collapse operator $\hat C=\sqrt{\kappa}\hat a$, and evolves according to the Lindblad master equation
\begin{equation}
    \frac{d\hat\rho}{dt}
    =
    -i[\hat H_{\rm int},\hat\rho]
    +
    \kappa\mathcal{D}[\hat a]\hat\rho,
    \label{eq:appendix_master_decay}
\end{equation}
where the interaction Hamiltonian is the same as in Eq.~\eqref{eq:Hint_main_ap} and the Lindbladian operator is
\begin{equation}
    \mathcal{D}[\hat a]\hat\rho
    =
    \hat a\hat\rho \hat a^\dagger
    -
    \frac{1}{2}
    \left\{
    \hat a^\dagger \hat a,\hat\rho
    \right\},
\end{equation}
Since the dispersive Hamiltonian is a function of the photon-number operator, it commutes with $\hat n$: $[\hat n,\hat H_{\rm int}]=0$. Thus, the QPE interaction changes only phases and does not modify the field energy.

For an arbitrary operator $\hat O$, Eq.~\eqref{eq:appendix_master_decay} gives
\begin{equation}
    \frac{d}{dt}\langle \hat O\rangle
    =
    i\langle[\hat H_{\rm int},\hat O]\rangle
    +
    \kappa
    \left\langle
    \hat a^\dagger \hat O \hat a
    -
    \frac{1}{2}
    \left\{
    \hat a^\dagger \hat a,\hat O
    \right\}
    \right\rangle .
    \label{eq:adjoint_lindblad}
\end{equation}
Choosing $\hat O=\hat n$, the Hamiltonian contribution vanishes and we obtain
\begin{align}
    \frac{d}{dt}\langle \hat n\rangle
    &=
    \kappa
    \left\langle
    \hat a^\dagger \hat n \hat a
    -
    \frac{1}{2}
    \left\{
    \hat n,\hat n
    \right\}
    \right\rangle .
\end{align}
Using
\begin{equation}
    \hat a^\dagger \hat n \hat a
    =
    \hat a^\dagger \hat a^\dagger \hat a \hat a
    =
    \hat n(\hat n-1),
\end{equation}
we find
\begin{align}
    \frac{d}{dt}\langle \hat n\rangle
    &=
    \kappa
    \left\langle
    \hat n(\hat n-1)-\hat n^2
    \right\rangle
    =
    -\kappa \langle \hat n\rangle .
    \label{eq:n_decay_rate}
\end{align}
Therefore, for an initial field with mean photon number $\langle \hat n(0)\rangle=N$,
\begin{equation}
    \langle \hat n(t)\rangle
    =
    N e^{-\kappa t}.
    \label{eq:n_decay_solution}
\end{equation}

Equivalently, one may introduce a semiclassical amplitude $\alpha_{\rm sc}(t)$ whose intensity reproduces the mean field energy, $|\alpha_{\rm sc}(t)|^2=\langle \hat n(t)\rangle_{\rm sc}$. The corresponding amplitude equation is
\begin{equation}
    \frac{d}{dt}\alpha_{\rm sc}
    =
    -\frac{\kappa}{2}\alpha_{\rm sc},
\end{equation}
with solution $\alpha_{\rm sc}(t)=\alpha_{\rm sc}(0)e^{-\kappa t/2}$. Choosing $|\alpha_{\rm sc}(0)|^2=N$ gives
\begin{equation}
    |\alpha_{\rm sc}(t)|^2
    =
    N e^{-\kappa t},
\end{equation}
which is the expression used as a benchmark in Fig.~\ref{fig:qpe_monitoring}. For the Fock-state simulations considered, $\alpha_{\rm sc}$ is only a classical intensity variable that reproduces the ensemble-averaged photon-number decay.

\section{Wigner reconstruction from displaced QPE readout}
\label{ap:wigner_qpe}

In this Appendix, we derive the relation between the Wigner function, displaced photon-number parity, and the QPE readout used in Sec.~\ref{sec:tomography}. We also define the Wigner-overlap fidelity used to quantify the numerical reconstructions.

\subsection{Displaced parity representation of the Wigner function}

We adopt the convention used in Ref.~\cite{lutterbach1997direct}, in which the Wigner function is written as
\begin{equation}
W(\alpha)
=
2\,{\rm Tr}
\left[
\rho\,D(\alpha)\Pi D^\dagger(\alpha)
\right],
\label{eq:Wigner_LD_convention}
\end{equation}
where the displaced, parity and number operator are, respectively, $D(\alpha)=\exp(\alpha a^\dagger-\alpha^*a)$, $\Pi=e^{i\pi \hat n}=(-1)^{\hat n}$ and $\hat n=a^\dagger a$.

Using the cyclic property of the trace and $D^\dagger(\alpha)=D(-\alpha)$, Eq.~\eqref{eq:Wigner_LD_convention} can be rewritten as
\begin{align}
W(\alpha)
&=
2\,{\rm Tr}
\left[
D^\dagger(\alpha)\rho D(\alpha)\Pi
\right]
\nonumber\\
&=
2\,{\rm Tr}
\left[
D(-\alpha)\rho D(\alpha)\Pi
\right].
\label{eq:Wigner_displaced_density}
\end{align}
Thus, if we define the displaced density operator
\begin{equation}
\rho_\alpha
\equiv
D(-\alpha)\rho D(\alpha),
\label{eq:rho_alpha_definition}
\end{equation}
then
\begin{equation}
W(\alpha)
=
2\,{\rm Tr}\left[\rho_\alpha \Pi\right].
\label{eq:Wigner_displaced_parity}
\end{equation}
Equivalently, if one defines the displaced state as $D(\alpha)\rho D^\dagger(\alpha)$, the same expression is obtained with the replacement $\alpha\rightarrow-\alpha$. This sign difference is purely conventional and corresponds to the choice of which displacement is applied before evaluating the parity.

Given that applying the parity operator to a Fock state yields $\Pi\ket{n}=(-1)^n\ket{n}$, and that expanding the trace in the Fock basis, we obtain
\begin{align}
W(\alpha)
&=
2
\sum_{n=0}^{\infty}
\bra{n}\rho_\alpha \Pi \ket{n}
\nonumber\\
&=
2
\sum_{n=0}^{\infty}
(-1)^n
\bra{n}\rho_\alpha\ket{n}.
\label{eq:Wigner_parity_sum_appendix}
\end{align}
The diagonal elements of $\rho_\alpha$ are precisely the photon-number probabilities of the displaced state,
\begin{equation}
P_n(\alpha)
=
\bra{n}
D(-\alpha)\rho D(\alpha)
\ket{n}.
\label{eq:Pn_displaced_definition}
\end{equation}
Therefore,
\begin{equation}
W(\alpha)
=
2\sum_{n=0}^{\infty}
(-1)^n P_n(\alpha).
\label{eq:Wigner_from_Pn_exact}
\end{equation}
Equation~\eqref{eq:Wigner_from_Pn_exact} is the basic relation used in the tomography protocol: once the photon-number distribution of the displaced field is known, the Wigner function follows from the parity-weighted sum of that distribution. This is the same physical principle underlying direct Wigner measurements based on displaced parity in cavity-QED and related platforms~\cite{lutterbach1997direct,bertet2002direct,deleglise2008reconstruction}.

\subsection{Connection with the QPE output}

If the photon-number support of the displaced state lies inside the resolved interval $0\leq n<N_{\max}$, then $P_y^{\rm QPE}(\alpha)=P_y(\alpha)$. In this resolved regime, Eq.~\eqref{eq:Wigner_from_Pn_exact} becomes
\begin{equation}
W_{\rm QPE}(\alpha)
=
2\sum_{y=0}^{N_{\max}-1}
(-1)^yP_y^{\rm QPE}(\alpha).
\label{eq:W_QPE_appendix}
\end{equation}

When the support of $P_n(\alpha)$ extends beyond the chosen Hilbert-space cutoff or beyond the resolved number interval, the reconstructed number distribution is no longer a faithful representation of the true displaced distribution. Although parity is insensitive to shifts by an even multiple of $N_{\max}$ in the ideal modular readout, a finite numerical or experimental implementation still requires a sufficiently large Hilbert-space window to represent the displaced state accurately.

\subsection{Shot-noise estimate for one phase-space point}

A single QPE shot at fixed $\alpha$ returns one register outcome $y_r$ in the $r$-th repetition. The corresponding single-shot estimator for the Wigner function is
\begin{equation}
w_r(\alpha)=2(-1)^{y_r}.
\label{eq:single_shot_wigner_estimator}
\end{equation}
After $N_{\rm shots}$ independent repetitions, the sample mean is
\begin{align}
\widehat W(\alpha)
&=
\frac{1}{N_{\rm shots}}
\sum_{r=1}^{N_{\rm shots}}
w_r(\alpha) \nonumber \\
&=
\frac{2}{N_{\rm shots}}
\sum_{r=1}^{N_{\rm shots}}
(-1)^{y_r}.
\label{eq:sample_mean_wigner_appendix}
\end{align}
The random variable $w_r$ takes the values $\pm 2$, and its expectation value is $W(\alpha)$. Therefore,
\begin{equation}
{\rm Var}[w_r]
=
\langle w_r^2\rangle-\langle w_r\rangle^2
=
4-W^2(\alpha).
\label{eq:single_shot_variance_appendix}
\end{equation}
The variance of the sample mean is then
\begin{equation}
{\rm Var}\!\left[\widehat W(\alpha)\right]
=
\frac{4-W^2(\alpha)}{N_{\rm shots}}
\leq
\frac{4}{N_{\rm shots}}.
\label{eq:wigner_sample_variance_appendix}
\end{equation}
This expression gives a simple estimate of the number of QPE repetitions required to reconstruct each phase-space point with a desired statistical precision.

\subsection{Wigner-overlap fidelity}

To quantify the agreement between the reconstructed and ideal Wigner functions, we use the Hilbert--Schmidt overlap between the corresponding density operators. The required phase-space identity follows directly from the Cahill--Glauber representation of density operators and quasiprobability distributions~\cite{CahillGlauber1969}. In the symmetric-ordering representation, the Wigner function of a density operator $\rho$ is
\begin{equation}
W_\rho(\alpha)
=
{\rm Tr}\!\left[\rho\,T(\alpha,0)\right],
\label{eq:Wigner_CG_definition}
\end{equation}
where, in the convention used throughout this work,
\begin{equation}
T(\alpha,0)
=
2D(\alpha)\Pi D^\dagger(\alpha).
\label{eq:T_alpha_zero}
\end{equation}
This is the same convention used in Eq.~\eqref{eq:Wigner_LD_convention}, for which $-2\leq W(\alpha)\leq 2$ and
\begin{equation}
{\rm Tr}(\rho)
=
\frac{1}{\pi}
\int d^2\alpha\, W_\rho(\alpha).
\label{eq:wigner_normalization_appendix}
\end{equation}

The Cahill--Glauber expansion states that a bounded operator $F$ can be represented as
\begin{equation}
F
=
\frac{1}{\pi}
\int d^2\alpha\,
f_F(\alpha)\,T(\alpha,0),
\label{eq:operator_expansion_CG}
\end{equation}
where its symmetric phase-space symbol is
\begin{equation}
f_F(\alpha)
=
{\rm Tr}\!\left[F\,T(\alpha,0)\right].
\label{eq:operator_symbol_CG}
\end{equation}
Taking the expectation value of $F$ in the state $\rho$ gives
\begin{align}
{\rm Tr}(\rho F)
&=
{\rm Tr}\!\left[
\rho\,
\frac{1}{\pi}
\int d^2\alpha\,
f_F(\alpha)\,T(\alpha,0)
\right]
\nonumber\\
&=
\frac{1}{\pi}
\int d^2\alpha\,
f_F(\alpha)
{\rm Tr}\!\left[\rho T(\alpha,0)\right]
\nonumber\\
&=
\frac{1}{\pi}
\int d^2\alpha\,
f_F(\alpha) W_\rho(\alpha).
\label{eq:expectation_CG}
\end{align}

Now choose $F=\rho_2$ and $\rho=\rho_1$. In this case, the phase-space symbol of $F$ is precisely the Wigner function of $\rho_2$:
\begin{equation}
f_{\rho_2}(\alpha)
=
{\rm Tr}\!\left[\rho_2 T(\alpha,0)\right]
=
W_{\rho_2}(\alpha).
\label{eq:symbol_density_is_wigner}
\end{equation}
Substituting Eq.~\eqref{eq:symbol_density_is_wigner} into Eq.~\eqref{eq:expectation_CG}, we obtain
\begin{equation}
{\rm Tr}(\rho_1\rho_2)
=
\frac{1}{\pi}
\int d^2\alpha\,
W_{\rho_1}(\alpha)W_{\rho_2}(\alpha).
\label{eq:HS_overlap_wigner}
\end{equation}
This is the phase-space form of the Hilbert--Schmidt inner product for the Wigner convention used in this work. 

In the main text, we apply Eq.~\eqref{eq:HS_overlap_wigner} with $\rho_1=\rho_{\rm id}$ and $\rho_2=\rho_{\rm QPE}$, defining
\begin{equation}
\mathcal F_W
=
\frac{1}{\pi}
\int d^2\alpha\,
W_{\rm id}(\alpha)W_{\rm QPE}(\alpha).
\label{eq:wigner_fidelity_appendix}
\end{equation}
For a pure target state, $\rho_{\rm id}
=
|\psi_{\rm id}\rangle\langle\psi_{\rm id}|$, this becomes
\begin{equation}
\mathcal F_W
=
{\rm Tr}(\rho_{\rm id}\rho_{\rm QPE})
=
\langle\psi_{\rm id}|\rho_{\rm QPE}|\psi_{\rm id}\rangle,
\label{eq:wigner_fidelity_pure_target}
\end{equation}
which is the usual state fidelity with respect to the ideal target state.

In the numerical implementation, the integral in Eq.~\eqref{eq:wigner_fidelity_appendix} is evaluated on the same phase-space grid used for the tomography:
\begin{equation}
\mathcal F_W
\simeq
\frac{1}{\pi}
\sum_{j,l}
W_{\rm id}(\alpha_{j,l})
W_{\rm QPE}(\alpha_{j,l})
\Delta x\,\Delta p .
\label{eq:wigner_fidelity_discrete}
\end{equation}
Because Eq.~\eqref{eq:wigner_fidelity_discrete} is evaluated on a finite grid and with a finite Hilbert-space cutoff, the resulting number is subject to numerical discretization errors, solver tolerances, finite-shot fluctuations, and truncation effects. In this work, we therefore use $\mathcal F_W$ as a benchmark of tomographic accuracy within the chosen numerical resources, rather than as an independent experimental certification of state preparation.

\bibliography{ref}

\end{document}